\documentclass[preprint,12pt]{elsarticle}

\usepackage{amssymb}
\usepackage{times}
\usepackage{graphics,latexsym}
\usepackage{graphicx}
\usepackage{amsmath}
\usepackage{tabularx}
\usepackage{color}

\newcommand{\be}{\begin{equation}}
\newcommand{\ee}{\end{equation}}
\newcommand{\bc}{\begin{center}}
\newcommand{\ec}{\end{center}}
\newcommand{\bdm}{\begin{displaymath}}
\newcommand{\edm}{\end{displaymath}}

\newcommand{\ds}{\displaystyle}

\journal{Mech. Res. Comm.}

\begin{document}

\begin{frontmatter}



\title{Effects of non-linear rheology on \\
the electrospinning process: a model  study}


\author[*]{Giuseppe Pontrelli\footnote{Corresponding author.}}
\author[*]{Daniele Gentili}
\author[*]{Ivan Coluzza}

\vspace{20mm} 

\author[**]{Dario Pisignano} 
\author[*]{Sauro Succi}

\address[*]{Istituto per le Applicazioni del Calcolo - CNR, Via dei Taurini 19-00185, Rome, Italy \\
tel. +39 0649270927, \;\; fax +39 064404306 \\
{\tt Email: giuseppe.pontrelli@gmail.com} \bigskip \\}
\address[**]{Dipartimento di Matematica e Fisica ``E. De Giorgi'', University of Salento \& National Nanotechnology Laboratory of Istituto Nanoscienze - CNR, Via Arnesano-73100 Lecce, Italy}
\begin{abstract}
We develop  an analytical bead-spring  model to investigate the role of
non-linear rheology on the dynamics of electrified jets
in the early stage of the electrospinning process.
Qualitative arguments, parameter studies as well as numerical simulations, show that
the elongation of the charged jet filament is significantly reduced in the presence of a non-zero yield stress.
This may have beneficial implications for the optimal design of future electrospinning experiments.

\end{abstract}


\end{frontmatter}

\section{Introduction}

The dynamics of charged polymers in external fields is an important problem in
non-equilibrium thermodynamics, with many applications in science and engineering \cite{doshi1995electrospinning, andra}.
In particular, such dynamics lies at the heart of electrospinning experiments, whereby 
charged polymer jets are electrospun to produce nanosized fibers; 
these are used for several applications, as reinforcing elements in 
composite materials, as building blocks of non-wetting surfaces layers on ordinary textiles, 
of very thin polymeric separation membranes, and of nanoelectronic and nanophotonic devices ~\cite{pisi, agar, arin, mann}.
In a typical electrospinning experiment, a charged polymer liquid is ejected at the nozzle 
and is accelerated  by an externally applied electrostatic field until it reaches down to a charged
plate, where the fibers are finally collected.
During the process, two different regimes take place: an initial stable phase,  
where the steady jet is accelerated by the field 
in a straight path away from the spinneret (the ejecting apparatus);  a
second stage, in which an electrostatic-driven bending instability arises before 
the jet reaches down to a collector (most often a grounded or biased plane), where the fibers are finally deposited.
In particular, any small disturbance, either a mechanical vibration at the nozzle 
or hydrodynamic perturbations within the experimental apparatus, 
misaligning the jet axis, would lead the jet into a
region of chaotic bending instability \cite{reneker2000bending}. 
The stretching of the electrically driven jet is thus governed by the competition 
between electrostatics and  fluid viscoelastic rheology.

The prime goal of electrospinning experiments is to minimize the radius of the collected 
fibers.  By a simple argument of mass conservation, this is tantamount to maximizing the 
jet length by the time it reaches the collecting plane. Consequently, the  bending
instability is a desirable effect, as long it can be kept under control in experiments. 
By the same argument, it is therefore of interest to minimize the length of the initial stable jet region. Analyzing such stable region
is also relevant for an effective comparison with results coming from electrospinning experiments studied in real-time by means of
high-speed cameras \cite{camp}  or X-ray phase-contrast imaging \cite{green}. \par
In the last years, with the upsurge of interest in nanotechnology, electrospinning has made the object of 
comprehensive studies, from both modelling~\cite{carroll2006electrospinning} and 
experimental viepoints ~\cite{theron2005multiple} (for a review see~\cite{carroll2008nanofibers}).
Two families of models have been developed: the first treats the jet filament as obeying the equations of continuum mechanics ~\cite{spivak2000model,feng2002stretching,feng2003stretching, hohman2001electrospinningI, hohman2001electrospinningII}.
Within the second one, the jet is viewed as a series 
of discrete elements
obeying the equations of Newtonian mechanics
~\cite{reneker2000bending, yarin2001bending}.
More precisely, the jet is regarded as a series of charged beads, connected by viscoelastic springs. 
Both approaches above typically assume Newtonian fluids, with a linear strain-stress
constitutive relation. 
On the other hand,  in a recent time, the use of viscoelastic fluids has 
also been investigated in a number of papers, both theoretical and experimental, 
for the case of power-law~\cite{feng2002stretching,spivak2000model} and
other viscoelastic fluids~\cite{carroll2006electrospinning,carroll2011discretized}, with special 
attention to the instability region. \par

In this paper, we investigate the effects of Herschel-Bulkley 
non-Newtonian rheology on the early stage of the jet dynamics.
The main finding is that the jet elongation during such initial 
stable phase can be considerably slowed down for the case of yield-stress fluids.
As a result, the use of yield-stress fluids might prove beneficial for the design of
future electrospinning experiments.

\section{The model problem} 

\setcounter{equation}{0}

\begin{figure}[ht!]
\centering\scalebox{0.3}{\includegraphics{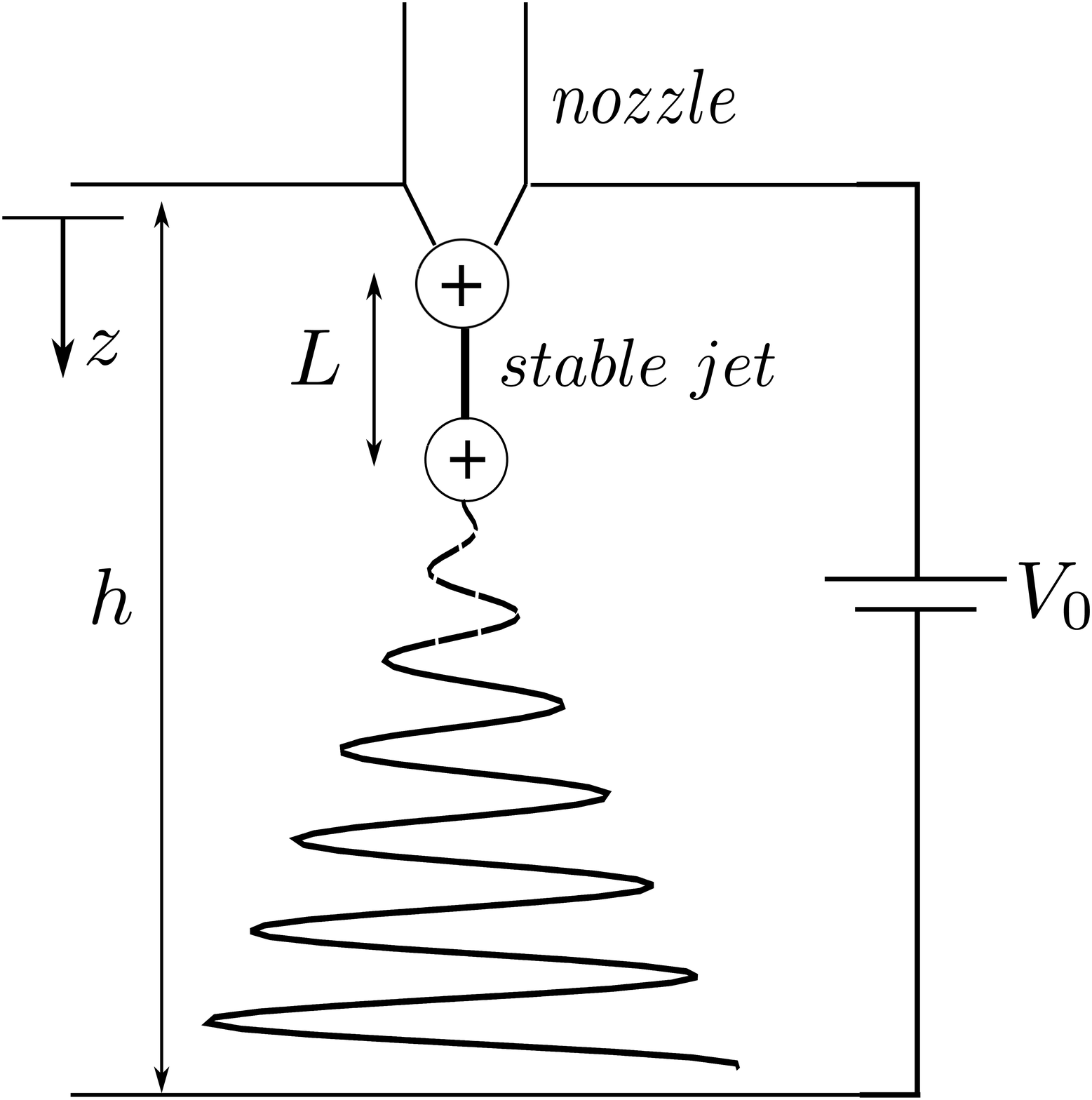}} 
\caption{The experimental set up and reference system of the stable jet region, with the origin at the nozzle orifice 
and $z$ coordinate axis pointing down (figure not to scale).
}
\label{fig:setup}
\end{figure}

Let us consider the electrical driven liquid jet in the electrospinning experiment.  
We confine our attention to the initial rectilinear stable jet region and, for simplicity, all variables 
are assumed to be uniform across the radial section of the jet, and vary along $z$ only, thus 
configuring a one-dimensional model.  
The filament is modelled by two charged beads ({\em dimer}) of mass $m$ and charge $e$,
separated by a distance $l$, and subjected to the external electrical 
field $V_0/h$, $h$
being the distance of the collector plate from the injection point (Fig.~\ref{fig:setup})
and $V_0$ the applied voltage.

The deformation of the fluid filament is governed by the combined action of
electrostatic and viscoelastic forces (gravity and surface tension are neglected), so that
the momentum equation reads \cite{reneker2000bending}:
\be
m {dv \over dt}=- {e^2 \over l^2} + {e V_0 \over h} + \pi a^2 \sigma\,,
\label{eqn1}
\ee
where $a$ is the cross-section radius of the bead and $v$ the velocity defined as:
\be
{dl \over dt}=-v
\label{eqn2}
\ee

For a viscoelastic fluid, the stress $\sigma$ is governed by the following equation:
\be
{d \sigma \over dt}=-\frac{1}{\tau}\left(\sigma-\sigma_{HB}\right)\,,
\label{eqn3}
\ee
where $\tau$ is the time relaxation constant and $\sigma_{HB}$ is the Herschel-Bulkley stress \cite{huang1998herschel,burgos1999determination} that reads
\be
\sigma_{HB}=\sigma_{Y}+ K \left(\frac{dl}{ldt}\right)^{n}\,
\label{eqHB}
\ee
 In the previous expression, $\sigma_{Y}$ is the yield stress
, $n$ is the power-law index and $\mu_0=K \left|\ds{1 \over l}  \ds{dl \over dt} \right|
^{n-1}$ is the effective viscosity with $K$ a prefactor having dimensions $g s^{n-2} cm^{-1}$; the case
$n=1$ and $\sigma_Y=0$ recovers the Maxwell fluid model, with $\mu_0 \equiv const.$ In the stress eqns. (\ref{eqn3})--(\ref{eqHB}), the Maxwell, the power-law and the 
Herschel-Bulkley models are combined.
A large class of polymeric and industrial fluids are described by  $\sigma_Y>0$ (Bingham fluid) 
and $n < 1$ (shear-thinning fluid), $n > 1$ (shear-thickening fluid) \cite{bird1987dynamics, pontrelli97, succi09}.  

It is expedient to recast the above eqns. in a nondimensional form
by defining a length scale and a reference stress as in \cite{reneker2000bending}: 
\be
L= \left({e^2 \over \pi a_0^2G}\right)^{1 \over 2}   \qquad \qquad G={\mu_0 \over \tau} 
\ee
with $a_0$ the initial radius. 
With no loss of generality, we assume  the initial length of the dimer to be $L$. 
Space is scaled in units of the equilibrium length $L$ at which
Coulomb repulsion matches the reference viscoelastic stress $G$,
while time is scaled with the relaxation time $\tau$.
The following nondimensional groups:

\be
Q= {e^2 \mu_0^2 \over L^3 m G^2}  \qquad\qquad V= {e V_0  \mu_0^2 \over h L m G^2} 
\qquad\qquad F= {\pi a_0^2 \mu_0^2 \over  L m G}
\ee

\noindent measure the relative strength of Coulomb, electrical, and viscoelastic forces respectively \cite{reneker2000bending}. Note that the above scaling implies $F=Q$. By setting  $W=-v$ and applying mass conservation:
\bdm
\pi a^2 l= \pi a_0^2 L
\edm
the above equations (\ref{eqn1})--(\ref{eqHB}) take the following nondimensional form: 
\begin{align}
\label{DYN} 
& {dl \over dt}= W \nonumber \\
& {dW \over dt}= V + {Q/l^2} - {F \sigma/l} \nonumber \\
& {d \sigma \over dt}= \sigma_Y+ ({W/l})^n - \sigma 
\end{align}
with initial conditions: $l(0)=1, W(0)=0, \sigma(0)=0$.
 Eqs. (\ref{DYN}) describe a dynamical system with 
non-linear dissipation for $n \ne 1$.
It can conveniently be pictured as a particle rolling down the
potential energy landscape $E(l) = Q/l - Vl$. 
Since the conservative potential is purely repulsive, the
time-asymptotic state of the system is escape to infinity, i.e.
$l \to \infty$ as $t \to \infty$.
However, because the system also experiences 
a non-linear dissipation, its transient dynamics is non-trivial.
This may become relevant to electrospinning experiments, as they take place in
set-up about and below 1 meter size, so that transient effects dominate the scene.

Before discussing numerical results, we firstly present a qualitative analysis of the problem.  

\section{Qualitative analysis}
\setcounter{equation}{0}
In the following we discuss some metastable and asymptotic 
regimes associated with the set of eqs. (\ref{DYN}) for $\sigma_Y=0$ and $n=1$ for simplicity, even though the qualitative conclusions apply to the general case as well (sect. 4).

\vspace{20mm} 
\underline{Accelerated expansion: free-fall}

In the absence of any Coulomb interaction and viscous drag
($Q=F=0$), the particle would experience a free-fall regime
\begin{equation}
\label{FREE}
l(t) = l_0 + W_0 t + Vt^2/2 = l_0 + Vt^2/2 \propto t^2
\end{equation}
for $t \gg 1$.
The same regime would be attained whenever Coulomb repulsion 
comes in exact balance with stress pullback, i.e.,
\begin{equation}
\sigma = \frac{Q}{Fl}= \frac{1}{l}
\end{equation}
Since $ \ds{d^2 l \over dt^2}= V$, one has $\sigma \to 1/t^2$ as $l \rightarrow \infty$, configuring again accelerated 
free-fall as the time-asymptotic regime of the system.  

\bigskip
\underline{Linear expansion}

Another possible scenario is the linear escape, i.e. $\ds{dW \over dt}=0$, yielding:
\begin{equation}
\label{LINESC}
l(t) \propto t
\end{equation}

This is obtained whenever the viscous drag exceeds over the Coulomb repulsion by just the
amount supplied by the external field $V$, namely: 
\begin{equation}
\label{SIGMA1}
\sigma = \frac{Q/l+Vl}{F}= \frac{1}{l} + \frac{V}{Q} l
\end{equation}
leaving $\ds{d^2 l \over dt^2}=0$.
This shows that, in order to sustain a linear growth, the stress 
should diverge linearly with the dimer elongation.
Again, this is incompatible with any asymptotic state of the stress evolution.
However, if $V$ is sufficiently small, namely:
\begin{equation}
l < l_Q = (Q/V)^{1/2}  \label{yhj}
\end{equation}
such regime may be realized on a transient basis.
Note that $l_Q$ designates the length below which Coulomb repulsion prevails
over the external field.

As we shall show, the solution $l \sim t$, $\sigma \sim 1/l \sim 1/t$ can indeed 
be attained as a transient quasi-steady state regime.
Typical experimental values are $Q/V \sim 10$, so that $l_Q \sim 3-10$
indicating  that such regime could indeed be attained in experiments 
with elongations $l< 1-3$ cm  (see section 4). 
Note that the value of $l_Q$ is independent of the rheological model, this latter affecting
however the time it takes to reach the condition $l=l_Q$. 
To analyze this issue, let us consider the steady-state limit of the stress equation
for a generic value of the exponent $n$, i.e.
\begin{equation}
\sigma = \left(\ds{1 \over l}\ds{dl \over dt} \right)^{n}
\end{equation}
The solution $l(t) \sim t$ delivers $\sigma \sim t^{-n} \sim l^{-n}$, which
is indeed compatible with the condition (\ref{SIGMA1}) for the case $n=1$.
Of course this is not an exact solution, since $\ds{d \sigma \over dt} = 0$ implies $\sigma=const$ in time .
However, it can be realized as a quasi-solution, in the sense that $\ds{1 \over \sigma}\ds{d \sigma \over dt} \ll 1$.  \\
The  above analysis is relevant because electrospinning experiments take place
under finite-size and finite-time non-equilibrium conditions, and it is 
therefore of great interest to understand the transition time between the two regimes.
In particular, the bending instability leading to three-dimensional helicoidal structures sets in
after an initial stage in which the polymer jet falls down in a linear configuration.
Since the goal of the experiment is to maximize the length $l$ of the polymer fiber
by the time it reaches the collector plate, it appears instrumental to
trigger the bending instability as soon as possible, so as minimize the elongation of the initial stable jet.
The present study is essentially a parametric analysis of this initial stage.

\section{Numerical results}
\setcounter{equation}{0}

We have integrated the system of eqs (\ref{DYN}) with 
a velocity-Verlet like time marching scheme:

\begin{eqnarray*}
\hat l = l + W \Delta t + a \frac{\Delta t^2}{2} \nonumber\\
\sigma_{HB} =\left(\frac{2  W}{\hat l + l} \right)^{n} +\sigma_Y \\
\hat \sigma = e^{-\Delta t} \;\sigma + (1-e^{-\Delta t}) \sigma_{HB} ;  \nonumber \\
\hat a = V-F {\hat \sigma \over \hat l} \ + {Q \over \hat l^2} \nonumber \\
\hat W = W + \left(\frac{\hat a +a}{2}\right) \Delta t \nonumber 
\end{eqnarray*}
with $\Delta t$ the time step and boundary conditions $l(0)=1, \, 
W(0)=\sigma(0)=0$.
Energy conservation has been checked and found to hold up to the sixth digit 
for simulations lasting up to $10^6$ time steps.
\bigskip\\
\underline{Reference results in Maxwell fluid}

\begin{figure}[ht!]
\centering\scalebox{0.5}{\includegraphics{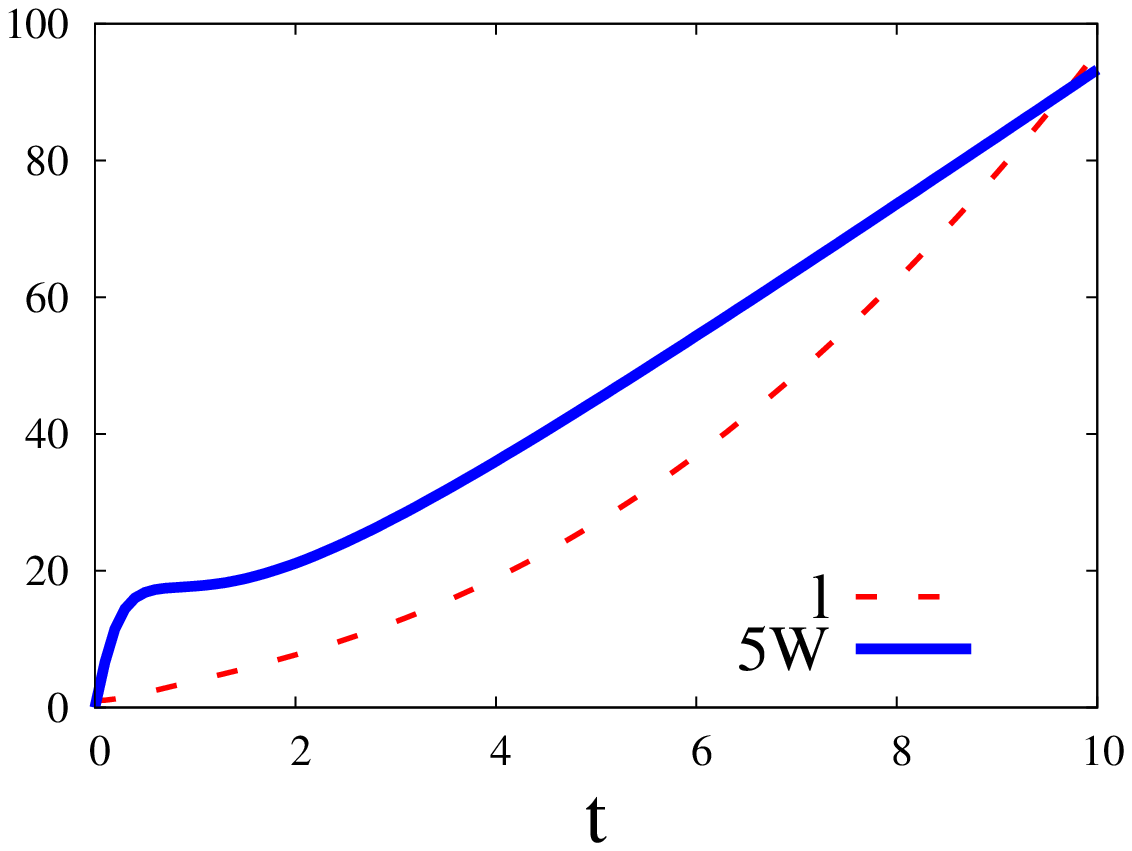}}  
\centering\scalebox{0.5}{\includegraphics{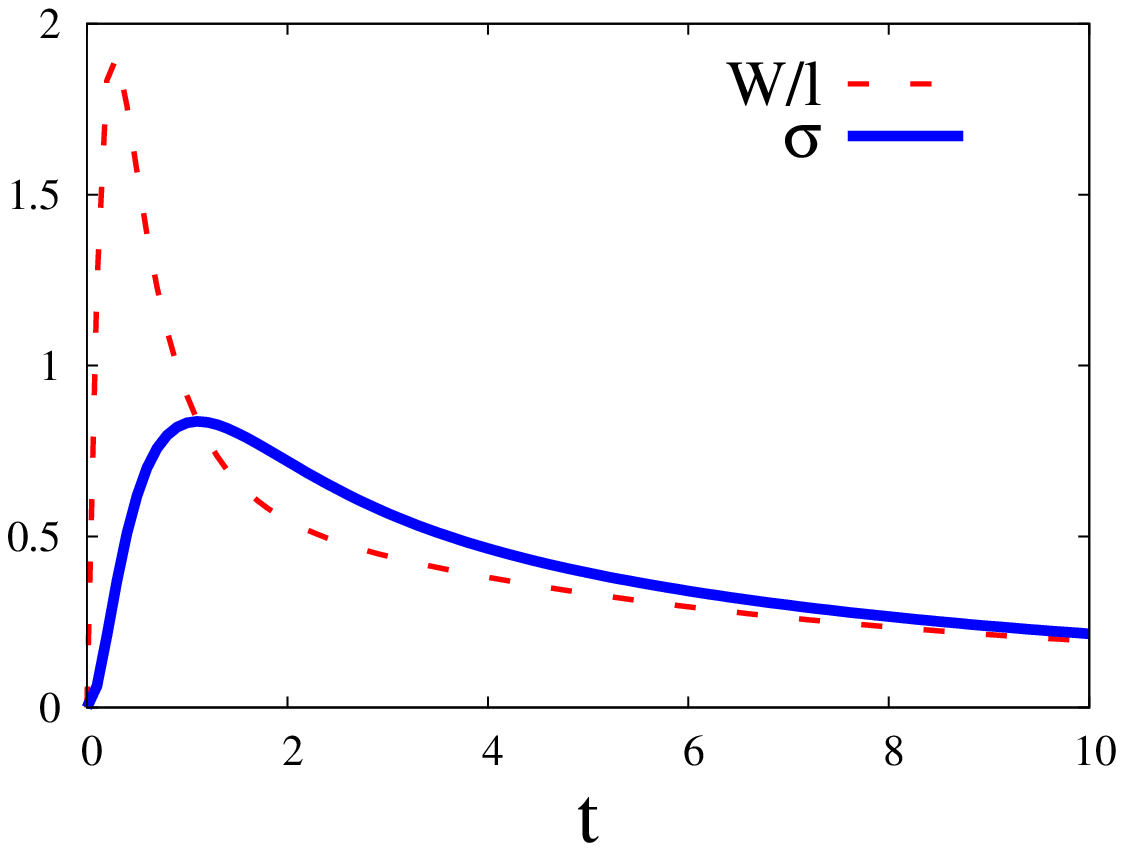}}   \caption{Time evolution of the elongation $l(t)$ and the velocity 
$5 \cdot W(t)$ (left), the stress $\sigma(t)$ and the strain rate $W/l$ (right) in the Maxwellian case $n=1, \sigma_Y=0$ (see text). 
}.   
\label{fig:newt}
\end{figure}

\begin{figure}[ht!]
\centering\scalebox{1.2}{\includegraphics{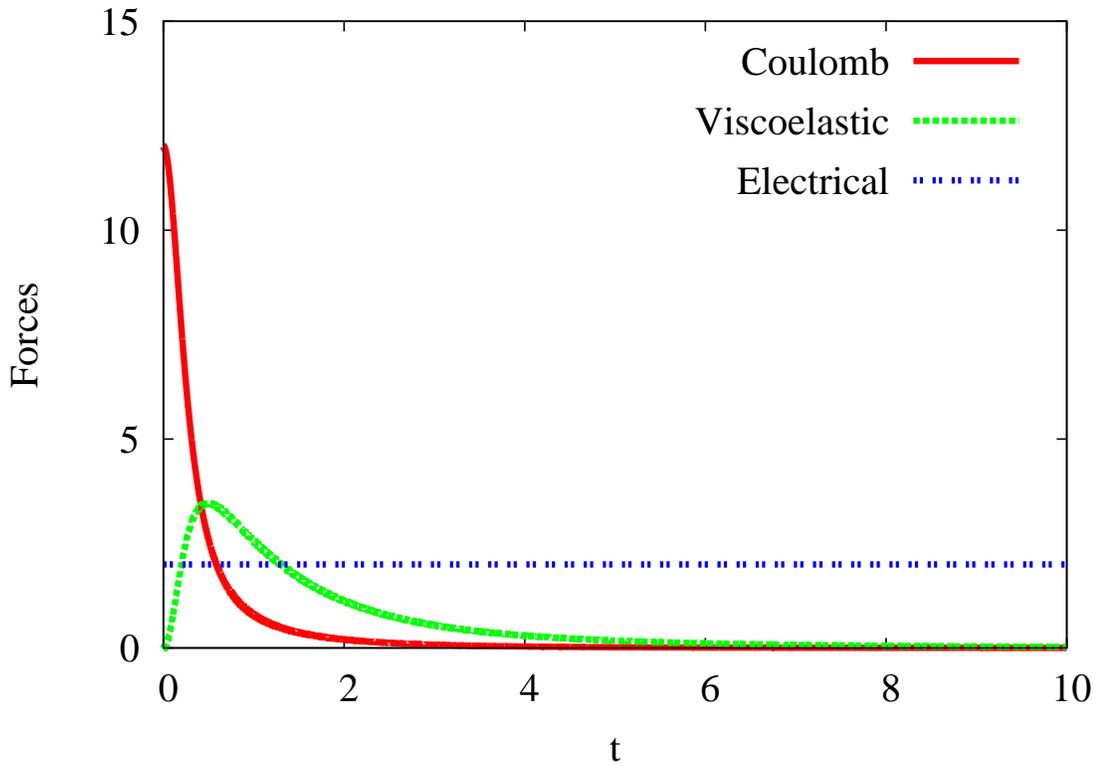}}  \\
\caption{The contribution of the repulsive Coulomb, the (opposite) viscoelastic and the electrical forces as a function of time on the system described by eqn. (\ref{DYN}) for $n=1$, $\sigma_Y=0$ (see text). 
}  
\label{fig:forces}
\end{figure}

As a reference case, we first consider the Maxwellian case $n=1, \sigma_Y=0$ 
and the typical values of experimental relevance are $L \sim 0.3$ cm, $\tau=10^{-2}$ s, yielding $Q=F=12$, $V=2$.
In fig. ~\ref{fig:newt} we report the time evolution of the elongation $l(t)$ and the velocity 
$W(t)$ (left), along with the stress $\sigma(t)$ and the strain rate $W/l$ (right). 
Three dynamic regimes are apparent.
First, an early transient, characterized by the build-up of velocity 
under the Coulomb drive and, to a much lesser extent, the external field as well.
As a result, the strain rate $W/l$ begins to grow, thus promoting a build-up of the stress,
which peaks at about $t=1.5$.
Subsequently, the stress starts to decay due to viscoelastic relaxation.
During the burst of the stress, lasting about up to $t=2$, the velocity comes
to nearly constant value, realizing the linear regime discussed in the previous section.
However, such regime cannot last long because the stress falls down very rapidly in time and
is no longer able of sustain the expanding "pressure" of the electrostatic interactions.
As clearly visible in figure \ref{fig:forces}, the Coulomb repulsion falls down faster than
the viscoelastic drag, and consequently the subsequent dynamics
is dominated by the external field, which promotes the quadratic scaling $l \sim t^2$, clearly
observed in Fig. \ref{fig:newt} at $t \gg 2$.
When both Coulomb repulsion and viscoelastic drag fall down to negligible values, which 
is seen to occur at about $t=5$ (50 ms in physical time), the free-fall regime sets in. 
At this time, the elongation has reached about $l=30$, corresponding to approximately 
$10$ cm in physical units. Taking $h=20$ cm as a reference value for the typical size of the
experiment, it is observed that the condition $l=h$ would be reached roughly at $t=8$, namely
$0.1$ s, corresponding to a mean velocity of about $2$ m/s, fairly comparable
with the experimental values. 

It is now of interest to explore to what extent such a picture is
affected by the fluid material properties. 
In particular, we wish to investigate whether
a non-Newtonian rheology is able to slow down the elongation dynamics, thereby reducing the stable jet length.
\bigskip \\
\underline{Effect of the shear-thinning and shear-thickening}

\begin{figure}[ht!]
\centering\scalebox{0.5}{\includegraphics{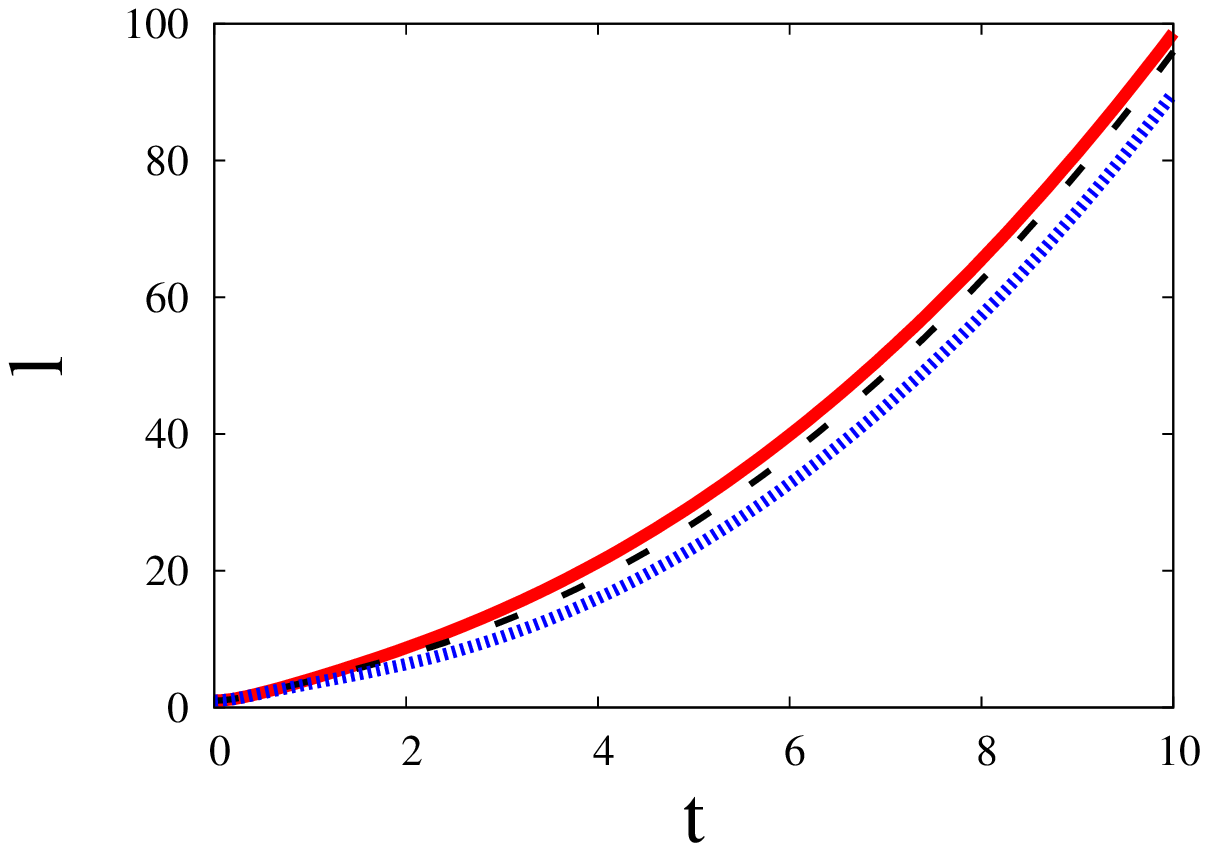}}  
\centering\scalebox{0.5}{\includegraphics{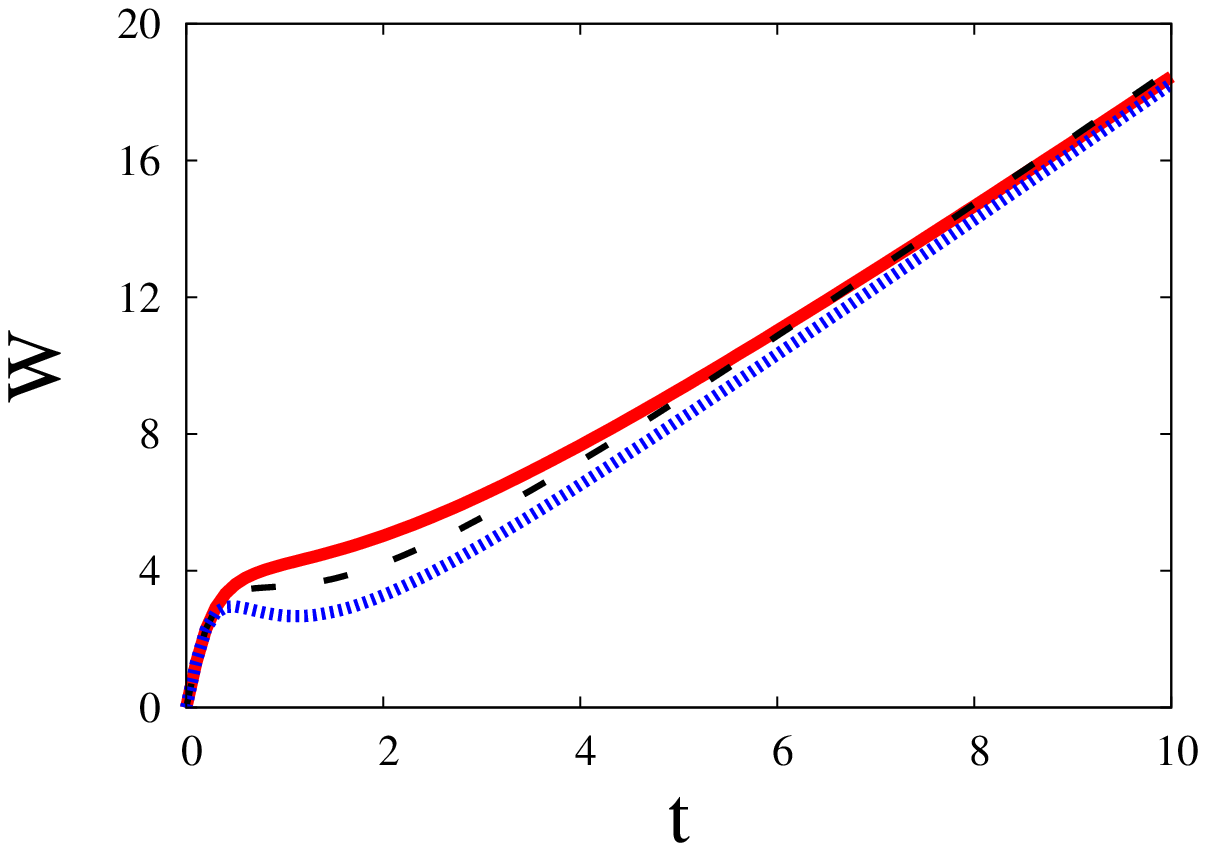}}  \\
\centering\scalebox{0.5}{\includegraphics{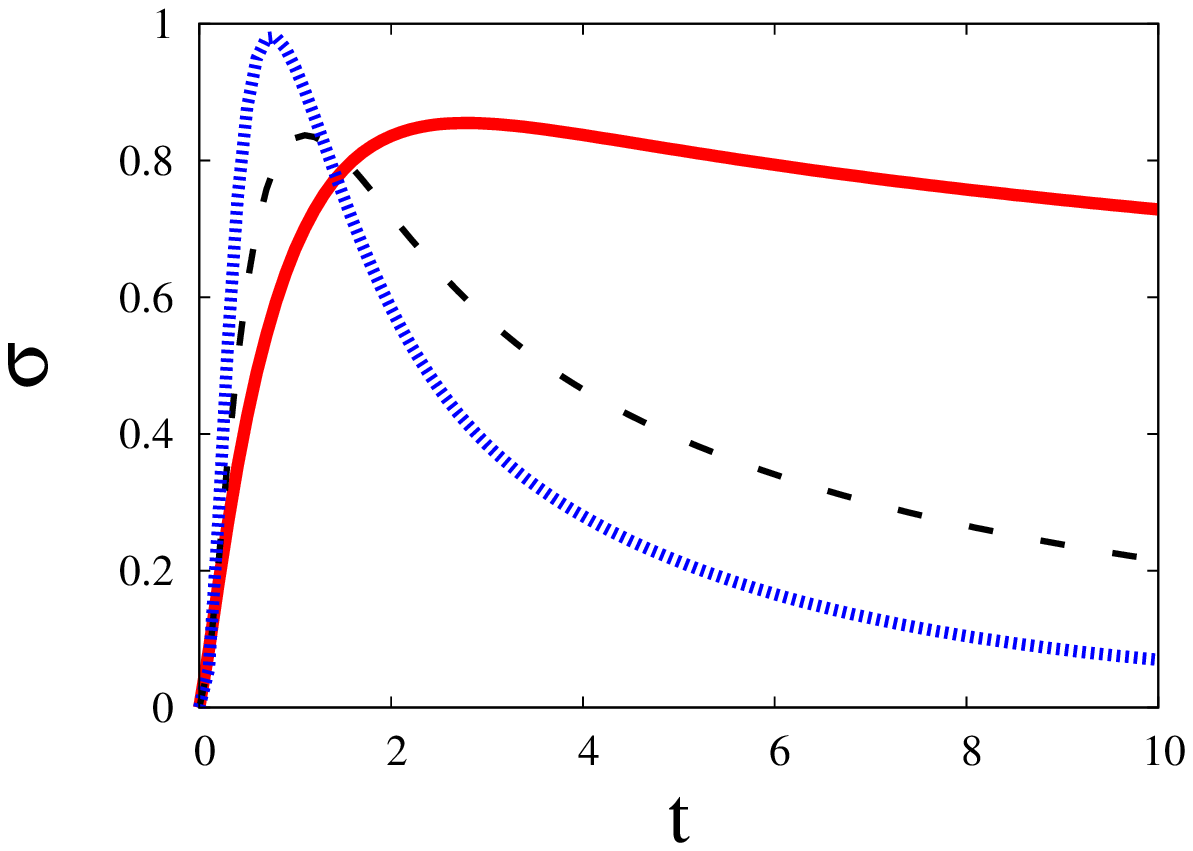}}
 \centering\scalebox{0.5}{\includegraphics{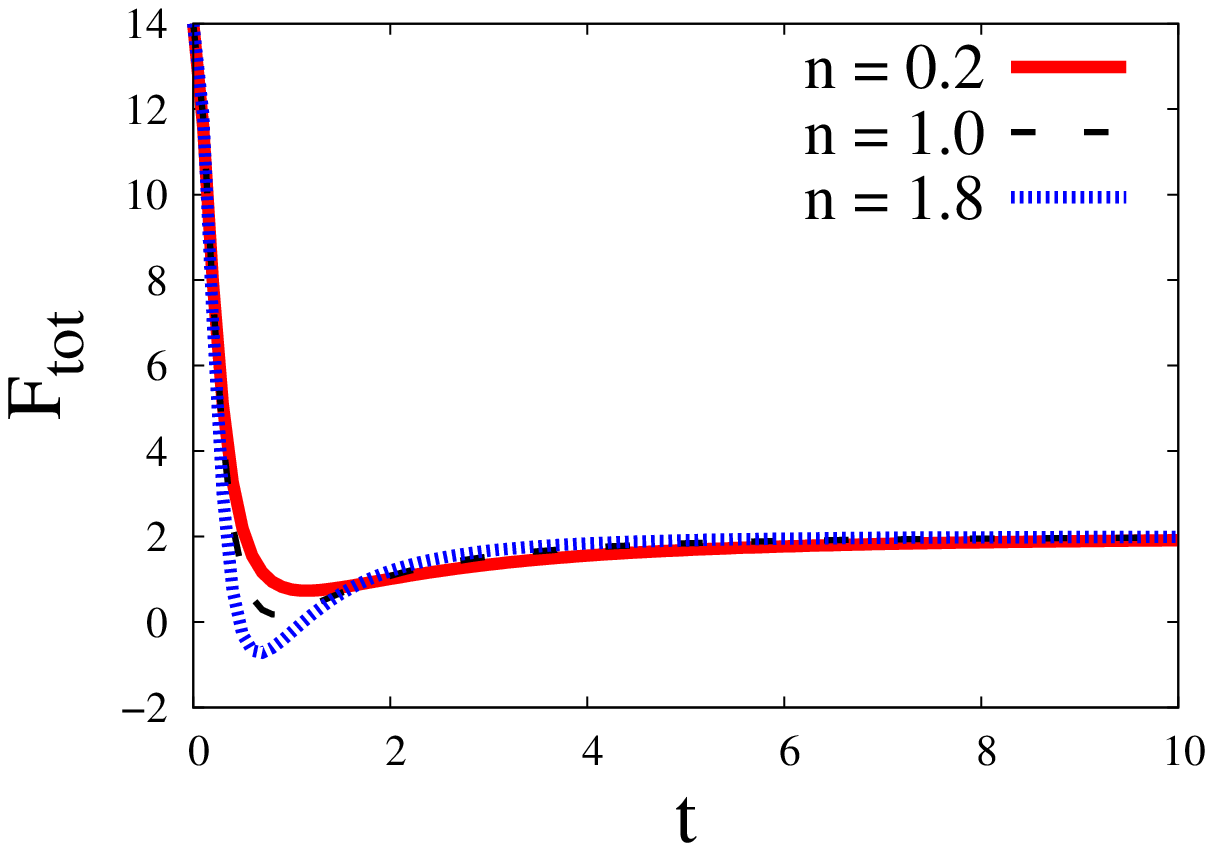}}  
\caption{The elongation, the velocity, the stress and the total force 
of the dimer as a function of time 
for three values of $n$ ($\sigma_Y=0$). 
For $n>1$, the dimer undergoes a shorter
elongation and velocity because in the transient $W/l>1$ and consequently the case
$n>1$ corresponds to a larger stress than $n=1$.}
\label{fig:beta}  
\end{figure}

To this purpose, the above simulations have been repeated for different values of 
$0.2< n < 1.8$, still keeping $\sigma_Y=0$.
In Fig. \ref{fig:beta}, we report $l,W,\sigma$ and the total force
$F_{tot}$ as a function of time, for $n=0.2,1,1.8$.
As one can see, the former case delivers the fastest growth.
This can be understood by noting that in the early stage of the evolution
$W/l>1$ (see Fig. ~\ref{fig:beta}), hence $n<1$ lowers (resp.  $n>1$ raises) the stress contribution 
as compared to the Maxwellian case, $n=1$.
To be noted that in the transient $1<t<2$, at $n=1.8$, the 
viscoelastic drag is able to produce a mildly decreasing velocity $W(t)$.
However, such mild decrease is very ephemeral, and is quickly replaced by a linear
growth like for the other values of $n$.
It is worth to note that for the case $n=0.2$, the stress remains 
substantial even at large times, which is reasonable because at $t \gg 1$ s $W/l \ll 1$.
However, the impact on the dimer elongation, $l$, is very mild, because the
factor $1/l$ makes the stress fairly negligible as compared to the external field. 
The final result is that the overall effect of $n$ on the dimer 
elongation is very mild, of the order of ten percent at most.
Hence the power-law model at zero yield-stress has 
a negligible effect on the jet length.
\bigskip\\
\underline{Influence of the yield stress} 

In the following, we investigate the effect of a non-zero yield stress, with fixed $n=1.8$ for convenience.
The condition $\sigma_Y>0$ is expected to slow down the growth of $l(t)$, because 
the stress decays to a non-zero value even in the infinite time limit.

\begin{figure}[ht!]
\centering\scalebox{0.5}{\includegraphics{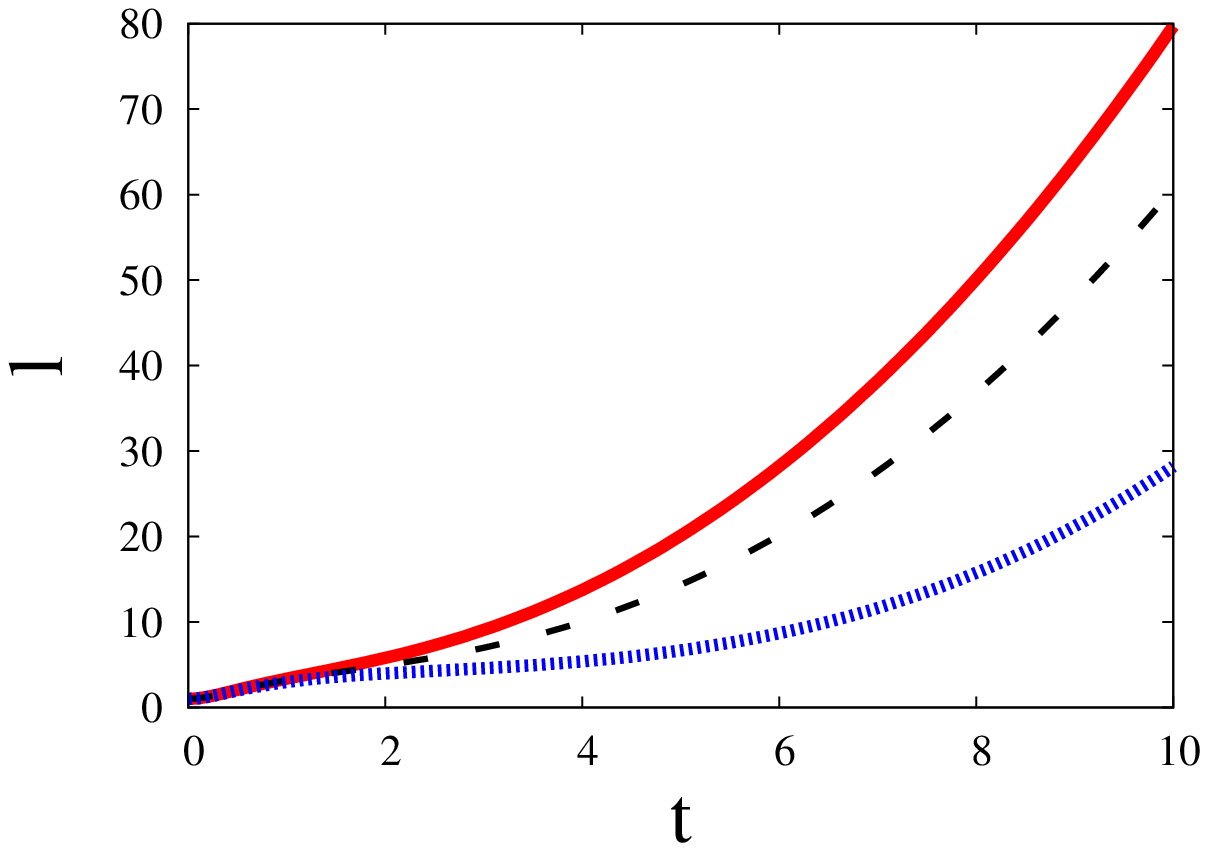}} 
\centering\scalebox{0.5}{\includegraphics{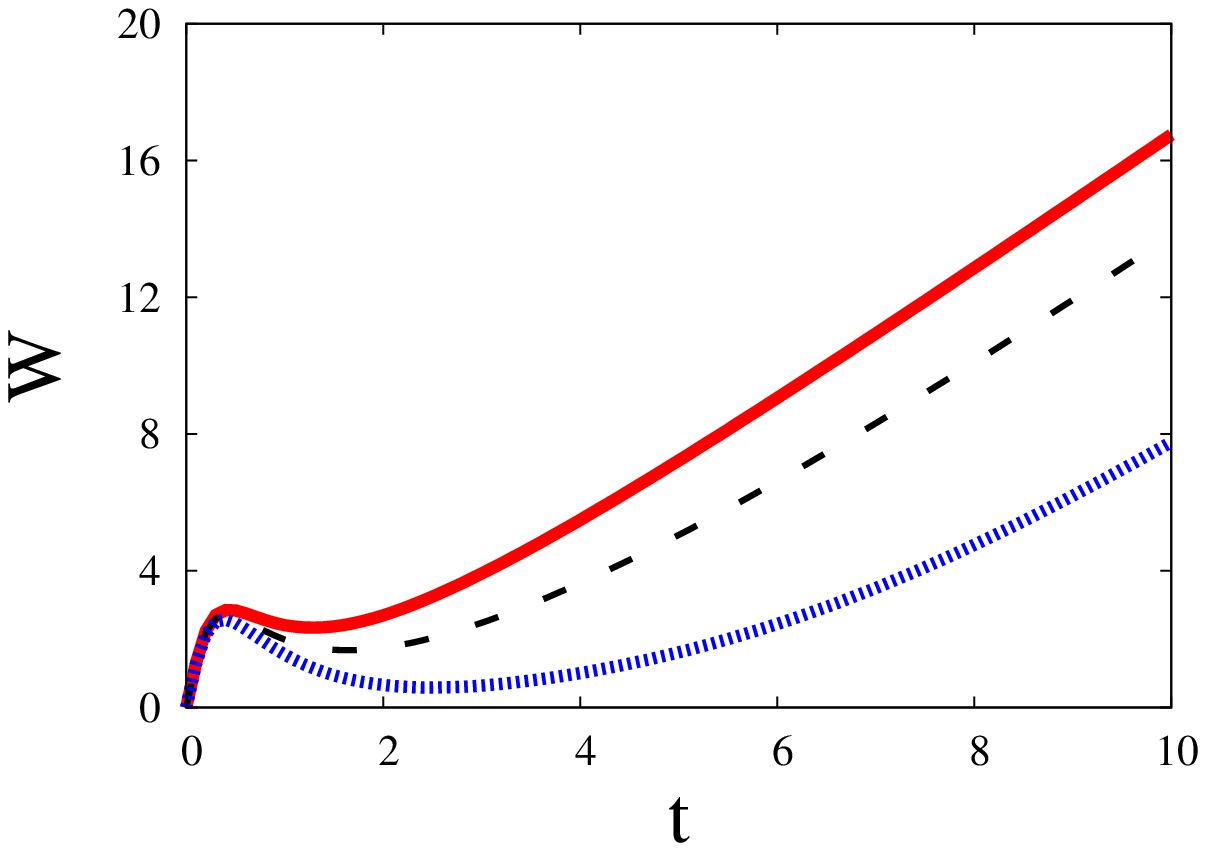}} \\
\centering\scalebox{0.5}{\includegraphics{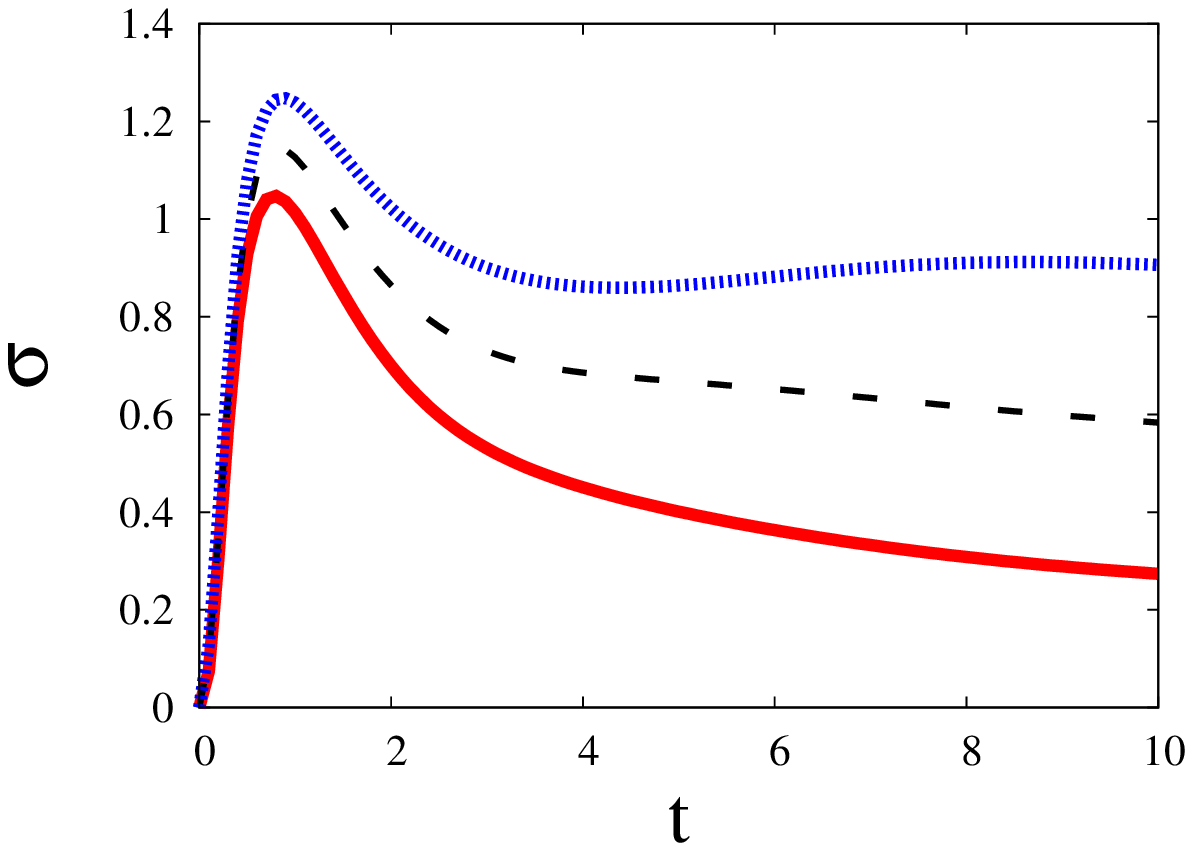}}
\centering\scalebox{0.5}{\includegraphics{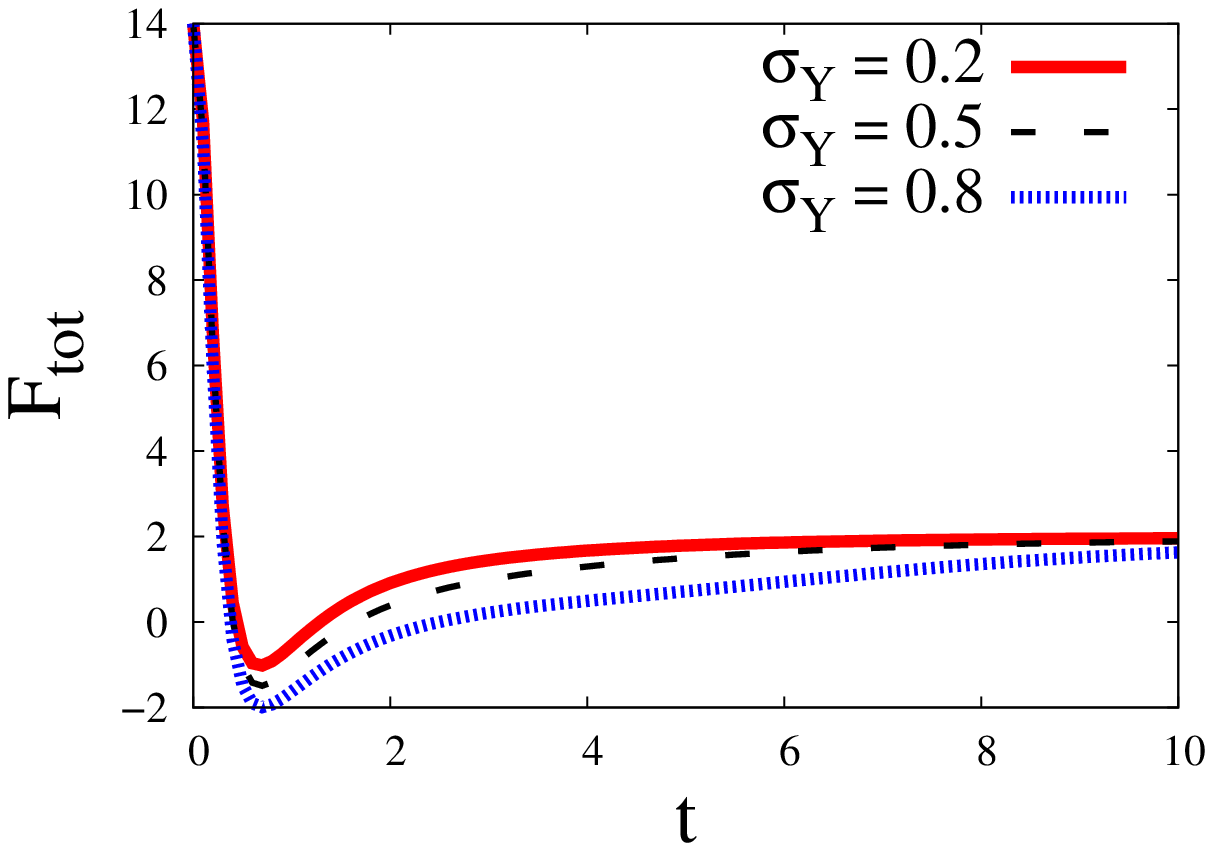}}

\caption{The effect of the yield stress $\sigma_Y$ on the elongation, the velocity, the stress and total forces ($n=1.8$);
Increasing the value of $\sigma_Y$ turns out to produce a significant reduction of the fluid elongation. } 
\label{fig:yield1} 
\end{figure}

Fig.~\ref{fig:yield1} shows the time evolution of $l,W,\sigma$ and  $F_{tot}$, for $\sigma_Y=0.2,0.5,0.8$. 
From this figure, it is readily appreciated that, at variance with the previous case,
increasing values of $\sigma_Y$ turn out to produce a significant slow down of the fluid elongation. 
In all case, the velocity $W(t)$ shows a decreasing trend in the transient $1<t<2$, coming very
close to zero at $t \sim 2.5$ for the case $\sigma_Y=0.8$.
The onset of the free-fall regime is significantly delayed, and consequently, so is
the evolution of the dimer elongation, which at $t=10$ reaches the values $l=80,60,30$
for $\sigma_Y=0.2,0.5,0.8$, respectively.
The latter case corresponds to a physical length of about $10 cm$, 
 about three times
shorter than in the Maxwell case, corresponding to about $l=90$ at $t=10$.
Hence, we conclude that yield stress fluids may experience a 
noticeable reduction of the stable jet length in actual electrospinning experiments.

\bigskip
\noindent \underline{Effective Forces}
 
Finally, it is of interest to inspect the effective force exerted upon 
the dimer as a function of its elongation. 
By effective force, we imply the sum of Coulomb repulsion and viscoleastic drag, namely
\begin{equation}
F_{eff}(l) = Q \left(\frac{1}{l^2} - \frac{\sigma(l)}{l}\right) 
\end{equation}
This expression may indeed provide useful input to coarse-grained models for 
three-dimensional simulations of the jet dynamics.
The effective force for different values of the exponent 
$n$ and yield-stress values $\sigma_Y$ is shown in Fig. 6, left and right panels.
From these figures, it is appreciated that the behavior of $F_{eff}$ as a function
of the elongation $l$ is similar to its dependence on time, although more peaked 
around the minimum. Such a minimum occurs slight above the crossover length,
$l_X$ at which Coulomb repulsion and viscoelastic drag come to an exact balance, i.e
\begin{equation}
l_X \sigma(l_X) = 1
\end{equation} 
For $l<l_X$ Coulomb repulsion is dominant, thus driving the stretching of the jet.
Subsequently, at $l>l_X$ the attractive component (drag) takes over, so 
that $F_{eff}(l)<0$, until a minimum is reached, $F_{min} \equiv F(l_{min})<0$.
Finally, for $l>l_{min}$, the force starts growing again to attain its asymptotic 
zero value at $l \to \infty$.
For the present choice of parameters, the minimum length $l_{min}$ is not far 
from the characteristic  length $l_Q=\left(\ds\frac{Q}{V}\right)^{1/2}$ (see eqn. (\ref{yhj})).
With the numerical values in point, $Q=12$ and $V=2$, 
we compute $l_Q \sim 2.44$.
Furthermore, such minimum length $l_{min}$ appears to be 
a decreasing function of $n$ at a given $\sigma_Y$ and independent 
of $\sigma_Y$ at a given $n$.

It is interesting to note that, upon rescaling the elongation with the 
computed values $l_{min}(n)$, the three curves with $n=0.2,1.0,1.8$
collapse to a universal function $F_{eff}(l) = A_n f(l/l_{min}(n))$, where $A_n$ is a
scaling amplitude which depends on the exponent $n$.
Such universal function might prove useful in the parametrization of three-dimensional interactions.

\section{Conclusions}

Summarizing, we have developed a model for the flow of electrically 
charged viscoelastic fluids, with the main aim of investigating the role of 
non-Newtonian rheology on the stretching properties of electrically charged jets.  
The simulations show good agreement with the theoretical 
analysis and provide a qualitative understanding of the role of viscoelasticity 
in the early stage of the electrospinning experiment. 
The main finding is that yield-fluids may lead to a 
significantly reduction of the linear extension of the jet in the initial
stage of the electrospinning process. 
The present findings may also prove useful to set up the model
parameters that control the efficiency of the process and the quality of the spun fibers. 
\bigskip\\

\underline{Acknowledgments} \\
The research leading to these results has received funding from the European Research Council under the European Union’s Seventh Framework Programme (FP/2007-2013)/ERC Grant Agreement n. 306357 (ERC Starting Grant “NANO-JETS”). 
One of the authors (S.S.) wishes to thank the Erwin Schroedinger Institute in Vienna, where this work
was initiated, for kind hospitality and financial support through the ESI Senior Fellow program.

\bibliographystyle{elsarticle-harv}
\bibliography{biblio}

\newpage

\begin{figure}[ht!]
\centering\scalebox{0.5}{\includegraphics{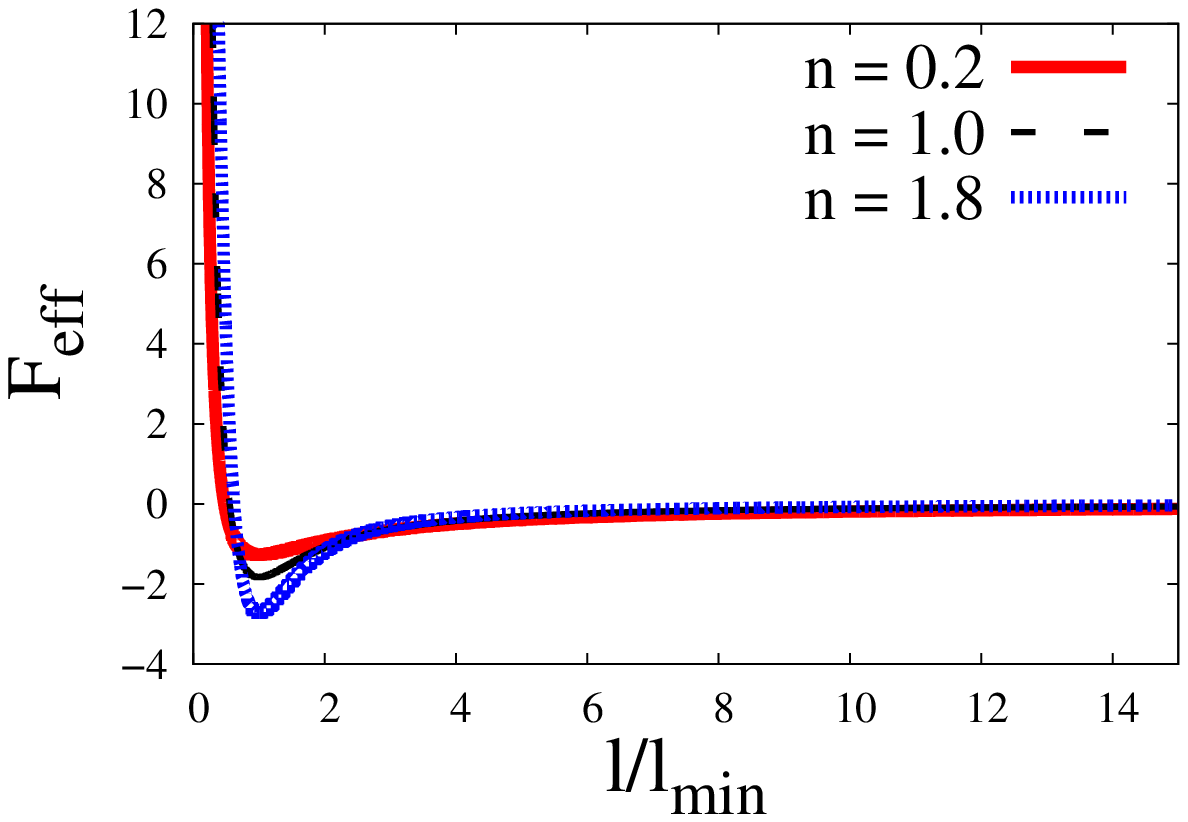}}
\centering\scalebox{0.5}{\includegraphics{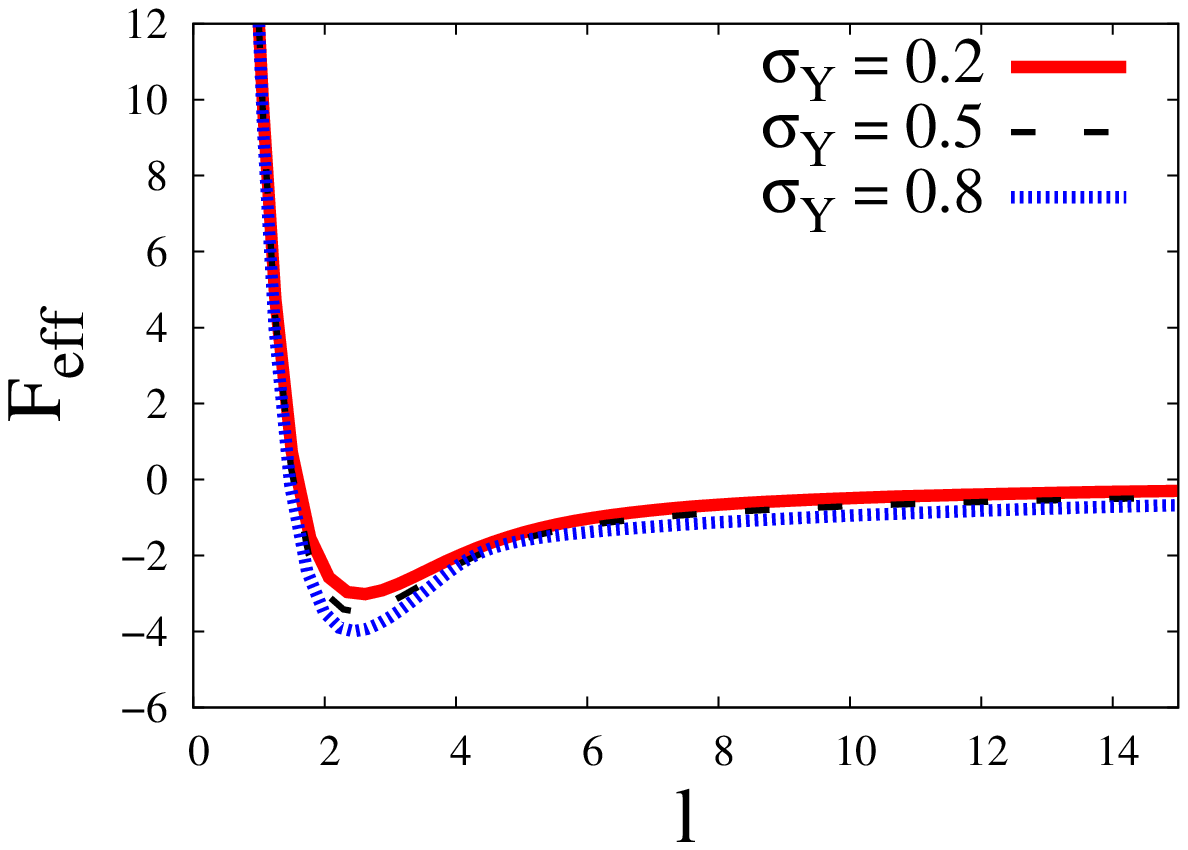}}
\caption{Effective forces as a function of the elongation $l$ for different
values of $n$ (left) at $\sigma_Y=0$, and for different $\sigma_Y$ at $n=1$ (right).
Note that in the left panel the elongation has been rescaled in units of $l_{min}$.} 
\label{fig:yield2} 
\end{figure}


\end{document}